\documentclass[a4paper,12pt]{article}

\usepackage{graphics}

\begin{document}

\title{
%\begin{flushright}
%{\normalsize IRB-TH-??/08}
%\end{flushright}
%\vspace{2 cm}
\bf Renormalization group scale-setting in astrophysical systems}

\author{Silvije Domazet\thanks{sdomazet@thphys.irb.hr} and Hrvoje \v Stefan\v ci\'c\thanks{shrvoje@thphys.irb.hr}}

%\author{Hrvoje \v Stefan\v ci\'c\thanks{shrvoje@thphys.irb.hr}}

\vspace{3 cm}
\date{
\centering
Theoretical Physics Division, Rudjer Bo\v{s}kovi\'{c} Institute, \\
   P.O.Box 180, HR-10002 Zagreb, Croatia}

%\institute{
%  Theoretical Physics Division, Ru\dj er Bo\v{s}kovi\'{c} Institute,
%   P. O. Box 1016, HR-10001 Zagreb, Croatia}

\maketitle

\abstract{A more general scale-setting procedure for General Relativity with Renormalization Group corrections is proposed. Theoretical aspects of the scale-setting procedure and the interpretation of the renormalization group running scale are discussed. The procedure is elaborated for several highly symmetric systems with matter in the form of an ideal fluid and for two models of running of the Newton coupling and the cosmological term. For a static spherically symmetric system with the matter obeying the polytropic equation of state the running scale-setting is performed analytically. The obtained result for the running scale matches the Ansatz introduced in a recent paper by Rodrigues, Letelier and Shapiro which provides an excellent explanation of rotation curves for a number of galaxies. A systematic explanation of the galaxy rotation curves using the scale-setting procedure introduced in this paper is identified as an important future goal.}

\vspace{2cm}

\section{Introduction}

The breakthrough in cosmological observations of the last two decades, followed by reshaping and refinement of cosmological models, has revealed a profound dichotomy of our present understanding of the universe. Expressed in terms of the (effective) energy budget of the universe, more that $95 \%$ of the universe consists of two mysterious dark components: dark matter and dark energy \cite{wmap}. We know what properties these two components, or their respective equivalent mechanisms, should have to reproduce the evolution of the universe that we observe. On the other hand, we have no decisive evidence on the physical nature of these components, or microscopical mechanisms that might lie behind them. 

Dark energy (DE) component of the universe is providing mechanism for the present phase of accelerated expansion of the universe \cite{de}. Its generic property is negative pressure needed to cause the acceleration in the cosmic expansion. There is a long list of candidates for the role of DE, including nondynamic or dynamic cosmological term, quintessence, phantom energy, etc. Present observational data are still consistent with many of these models including the benchmark $\Lambda$CDM model. Alternative mechanisms of accelerated expansion include e.g. modified gravity models \cite{modgrav} and braneworld models \cite{brane}.

Dark matter (DM) component accounts for almost a quarter of energy density of the universe ($23 \%$ in $\Lambda$CDM model \cite{wmap}). It is presently nonrelativistic and it has a key role in the growth of inhomogeneities and formation of structures that we presently observe in the universe \cite{padm}. Its influence is also felt at astrophysical scales. The presence of large quantities of dark matter is required to explain the galaxy rotation  curves and galactic cluster dynamics (for a recent review see e.g. \cite{DMref}). There abound particle candidates for dark matter \cite{DMref}, but alternative mechanisms such as MOND \cite{mond} and STVG \cite{stvg} have also been proposed.

Within the trend of introducing ad hoc descriptions of unexplained observed phenomena such as DE and DM, it is very useful to reconsider whether we could explain, or at least add to understanding of dark matter and dark energy in the well known physical framework.
The study of renormalization group running effects of Quantum Field Theory in curved space-time is one of such frameworks \cite{rungen1,rungen2}. This approach still raises considerable theoretical interest \cite{foot,rungen1} and it abounds in phenomenologically interesting applications. Indeed, the cosmology with running parameters from QFT in curved space-time was found illuminating in understanding of general evolution of the universe \cite{exp1,exp2,exp3,exp4}, how variable cosmic parameters could  mimic other forms of dark energy and related phenomena \cite{mimic}, the coincidence problem \cite{coin}, growth of inhomogeneities \cite{inhom} and even the relaxation mechanism \cite{mojrelax} for the cosmological term \cite{relax}.

In a recent paper \cite{galrot}, an approach to galaxy rotation curves based on General Relativity with Renormalization Group corrections (RGGR), previously initiated in \cite{mijcap}, was elaborated.  The authors of \cite{galrot} start from an Ansatz for the RG scale relevant for astrophysical objects, 
\begin{equation}
 \label{eq:ansatz}
\frac{\mu}{\mu_0}=\left( \frac{\phi}{\phi_0} \right)^{\alpha} \, ,
\end{equation}
where $\phi$ is the Newtonian gravitational potential and the subscript $0$ refers to some benchmark scale.  Using this Ansatz the galaxy rotation curves of nine galaxies are reproduced with impressive precision. The analysis of observational data on galaxy rotation curves showed that RGGR explained the data equally well as the isothermal sphere dark matter model and better than MOND and STVG models.

The most arbitrary part of the analysis given in \cite{galrot} is the ad hoc character of the Ansatz (\ref{eq:ansatz}) for the RG scale. The main aim of this paper is to introduce a systematic procedure of defining the scale of RGGR in astrophysical systems. Using this procedure we perform an analysis for spherically symmetric systems which provides  arguments in favor of the Ansatz (\ref{eq:ansatz}).

The presentation of the paper is organized in the following way. The first section is the introduction. In the second section the scale-setting procedure is introduced and elaborated. This section also contains the discussion of the interpretation of the scale and the presentation of physically interesting examples of scale-setting. The third section gives a detailed analysis of the scale-setting in spherically symmetric systems. The fourth section contains the discussion of the results and the fifth section ends the paper with the conclusions. 

\section{The scale-setting procedure}

\subsection{Formalism}

The classical description of gravity in General Relativity is based on the Hilbert-Einstein action \footnote{Our conventions are the following: the signature of the metric is $(+,-,-,-)$, the Christoffel symbols are $\Gamma_{\alpha \beta}^{\mu}=\frac{1}{2} g^{\mu \nu} (g_{\alpha \nu,\beta}+g_{\beta \nu,\alpha}-g_{\alpha \beta,\nu})$, the Riemann tensor is defined as $R^{\alpha}_{\;\; \eta \beta \gamma}=\Gamma^{\alpha}_{\beta \eta, \gamma}-\Gamma^{\alpha}_{\eta \gamma, \beta}+\Gamma^{\alpha}_{\tau \gamma} \Gamma^{\tau}_{\beta \eta}- \Gamma^{\alpha}_{\tau \beta} \Gamma^{\tau}_{\gamma \eta}$ and the Ricci tensor is $R_{\eta \gamma}=R^{\alpha}_{\;\; \eta \alpha \gamma}$. The symbol $,\alpha$ denotes an ordinary partial derivative with respect to coordinate $x^{\alpha}$.}
\begin{equation}
 \label{eq:GRclass}
S_{EH}=\frac{1}{16 \pi G} \int d^4 x \sqrt{-g} (R - 2\Lambda) \, ,
\end{equation}
where $\Lambda \equiv 8 \pi G \rho_{\Lambda}$. The action functional including matter $S=S_{EH}+S_{matter}$  under variation of metric $g_{\mu \nu}$ leads to Einstein equations
\begin{equation}
 \label{eq:Eineq}
G_{\alpha\beta} \equiv R_{\alpha\beta} - \frac{1}{2} R g_{\alpha\beta} = -8 \pi G (T_{\alpha\beta}+\rho_{\Lambda} g_{\alpha\beta}) \, ,
\end{equation}
where $T_{\alpha\beta}$ is the matter energy-momentum tensor.

Introduction of the renormalization group corrections to General Relativity can be performed at three distinct levels \cite{reuter}. The first way of introducing RG correction is at the level of solutions of Einstein equations. We may consider analytic solutions of classical Einstein equations (\ref{eq:Eineq}) for which the dependence of the solution on parameters such as $G$ and $\rho_{\Lambda}$ has a well-defined analytical form. We introduce RG corrections so that in the solutions we substitute ``classical'' parameters $G$ and $\rho_{\Lambda}$ with the RG running ones $G(\mu)$ and $\rho_{\Lambda}(\mu)$. This way of introducing the RG corrections is somewhat limited by the availability of analytical solutions in General Relativity. The second way is the introduction of the RG correction at the level of Einstein equations. In this approach the ``classical'' parameters $G$ and $\rho_{\Lambda}$ in Einstein equations (\ref{eq:Eineq}) are substituted with $G(\mu)$ and $\rho_{\Lambda}(\mu)$. The third way of introducing the RG corrections is at the level of action. The parameters $G$ and $\rho_{\Lambda}$ in (\ref{eq:GRclass}) are replaced with $G(\mu)$ and $\rho_{\Lambda}(\mu)$ \cite{reuter2}.

In this paper we consider the introduction of RG corrections at the level of Einstein equations. The RG corrected Einstein equations then read
\begin{equation}
 \label{eq:EineqRG}
G_{\alpha\beta} \equiv R_{\alpha\beta} - \frac{1}{2} R g_{\alpha\beta} = -8 \pi G(\mu) (T_{\alpha\beta}+\rho_{\Lambda}(\mu) g_{\alpha\beta}) \, .
\end{equation}
The RG scale $\mu$ present in (\ref{eq:EineqRG}) is a scalar under general coordinate transformations. Its physical meaning within the scale-setting procedure will be discussed in subsection \ref{mean}.

We assume that the matter component, with the energy-momentum tensor $T_{\alpha\beta}$ serving as a source in (\ref{eq:Eineq}), can be qualified as an ideal fluid. Then its energy-momentum tensor has the following form
\begin{equation}
 \label{eq:tensmatt}
T^{\alpha\beta} = (\rho+p) u^{\alpha} u^{\beta} - p g^{\alpha \beta} \, .
\end{equation}
Here $u^{\alpha}$ denotes the four-vector of velocity of the fluid. The matter energy-momentum tensor is covariantly conserved
\begin{equation}
 \label{eq:concons}
\nabla_{\alpha} T^{\alpha\beta} = 0 \, .
\end{equation}
Using the covariant conservation of the Einstein tensor, $\nabla_{\alpha} G^{\alpha \beta}=0$, from (\ref{eq:Eineq}) one readily obtains
\begin{equation}
 \label{eq:cons1}
\nabla_{\alpha}(G(\mu)(T^{\alpha\beta}+\rho_{\Lambda}(\mu) g^{\alpha\beta})) = 0 \, .
\end{equation}
Taking into account that $\mu$ is a scalar, Eq. (\ref{eq:cons1}) transforms to
\begin{equation}
 \label{eq:cons2}
(\partial_{\alpha} \mu) \left[ G'(\mu) (T^{\alpha\beta} + \rho_{\Lambda}(\mu) g^{\alpha \beta}) + G(\mu) \rho_{\Lambda}'(\mu) g^{\alpha\beta} \right] = 0 \, ,
\end{equation}
where primes denote differentiation with respect to $\mu$.
This expression gives the scale-setting condition for a general matter component in the RGGR theory.

For the matter component having the matter property of an ideal fluid, the scale-setting condition further transforms to
\begin{equation}
\label{eq:cons3} 
(\partial_{\alpha} \mu) \left[ G'(\mu) ((\rho+p)u^{\alpha} u^{\beta} + (\rho_{\Lambda}(\mu)-p) g^{\alpha \beta}) + G(\mu) \rho_{\Lambda}'(\mu) g^{\alpha\beta} \right] = 0 \, .
\end{equation}
Multiplying this expression with $u_{\beta}$ and contracting $\beta$ indices we finally obtain a useful relation
\begin{equation}
 \label{eq:cons4}
(u^{\alpha} \partial_{\alpha} \mu) \left[ G'(\mu)(\rho+\rho_{\Lambda}(\mu))+G(\mu) \rho_{\Lambda}'(\mu) \right ] = 0 \, .
\end{equation}

\subsection{Interpretation of the running scale}

\label{mean}

The identification of the RG running scale $\mu$ has recently received a lot of attention in the literature \cite{exp1,exp2,exp3,exp4}. The main argumentation on which these attempts are based is a notion that the RG scale should have a clear physical interpretation \footnote{As opposed to the renormalization scale which is a purely formal quantity devoid of physical meaning.}. Although several physically  motivated approaches to the definition of the RG scale might be adopted, here we just take into account the general covariance and treat the RG scale as a scalar quantity. 

In our approach in RGGR models we have two important elements related to the scale $\mu$: 
\begin{enumerate}
\item The running laws for Newton coupling and the cosmological constant (CC energy density which result in running quantities $G(\mu)$ and $\rho_{\Lambda}(\mu)$, 
\item  the identification of the RG scale $\mu$, i.e. its expression in terms of relevant physical quantities. 
\end{enumerate}
These elements are not independent as it is evident from (\ref{eq:cons2}).  Independent functional forms for $G(\mu)$ and $\rho_{\Lambda}(\mu)$ and an arbitrary choice for $\mu$ are in general not consistent with (\ref{eq:cons2}). This constraint can be accommodated in two different ways \footnote{If one allows for non-conservation of matter energy-momentum tensor, then both functional forms of $G(\mu)$ and $\rho_{\Lambda}(\mu)$ and an arbitrary choice for $\mu$ can can coexist at an expense of energy-momentum interchange with matter.}. In the first one, one selects a physically motivated scale $\mu$ and decides that one of the running laws for $G(\mu)$ and $\rho_{\Lambda}(\mu)$ is better motivated than the other. Finally, from the scale-setting condition (\ref{eq:cons2}) and Einstein equations the dynamics of the system is determined and the remaining running law is reconstructed, see e.g. \cite{mijcap}. Alternatively, one may take the running laws as derived from a  fundamental theory and thus more reliable and then deduce the RG scale $\mu$ from the scale-setting condition (\ref{eq:cons2}). The physical meaning of the calculated scale is then determined {\it a posteriori}. The latter approach has already been applied in cosmological context in \cite{miprd}. In this paper we extend it to a broader astrophysical domain.

The scale-setting procedure can be carried out analytically only for systems posessing some symmetry, as demonstrated in subsection \ref{examples}. The said symmetry is usually evident only in a specific coordinate system and the scale-setting procedure is carried out using these very coordinates. As a result, we obtain $\mu$ as a function of coordinates consistent with the symmetries. This does not mean that for $\mu$ we choose some specific {\it ad hoc} combination of coordinates which would be at odds with covariance, but that we have a scalar function $\mu$ the dependence of which on coordinates is determined in a well-defined scale-setting procedure.
The obtained functional dependence on coordinates is specific to the chosen coordinate system and it is not an Ansatz for all coordinate systems. A special challenge of the scale-setting procedure is, after we obtain its concrete form in specific coordinates, to translate the dependence of $\mu$ on coordinates (or metric) into dependence on other scalar quantities (such as e.g. $R$).   

\subsection{Running models}

A concrete form of the running laws proceeds from some fundamental theory such as Quantum Field Theory in curved space-time or Quantum Gravity. In this paper we focus of on two specific models of which one comes from the framework of Quantum Field Theory in curved space-time and the other is motivated from Quantum Gravity approach.

The QFT in curved space-time served as a theoretical framework for a model of the cosmological constant running introduced in \cite{exp3}. The assumption of the model is that the running law for the CC follows a so called soft decoupling running law $\rho_{\Lambda}(\mu)=c_0 + c_2 \mu^2 + \dots$, with $c_2=3 \nu M_{P}^2/4 \pi$ where $\nu$ is a parameter \cite{exp3}. Further arguing that $\mu=H$, where $H$ is the Hubble parameter, is a physically well motivated choice for the running scale in the cosmological setting, the running law for the Newton coupling follows:
\begin{equation}
 \label{eq:runG}
G(H,\nu)=\frac{G_0}{1+2 \nu \ln \frac{H^2}{H_0^2} } \, .
\end{equation}
Therefore, the above expression is obtained as the result of the running law for $\rho_{\Lambda}$ and identification $\mu=H$.
The assumption of validity of the functional form (\ref{eq:runG}) in general, i.e. not just for cosmology and $\mu=H$ is a central hypothesis of the authors of \cite{galrot} starting from which they build their explanation of galaxy rotation curves. In this paper the scale-setting procedure is applied to the model
\begin{equation}
 \label{eq:mod1rho}
\rho_{\Lambda}(\mu)=c_0 + c_2 \mu^2 \, 
\end{equation}
and 
\begin{equation}
 \label{eq:mod1G}
G(\mu)= \frac{G_0}{1+d_2 \ln \frac{\mu^2}{\mu_0^2} }\, .
\end{equation}
It is important to stress that here in general we consider $c_2$ and $d_2$ as independent parameters.

The running model proposed in \cite{bonnano} by Bonanno and Reuter postulates the existence of a nontrivial IR fixed point in the running of $G$ and $\Lambda$.
In this paper we consider a generalization of the running laws used in \cite{bonnano}. The running law for the cosmological constant energy density is 
\begin{equation}
 \label{eq:mod2rho}
\Lambda(k) \equiv 8 \pi G(k) \rho_{\Lambda}(k) = A k^{\alpha}
\end{equation}
 and the corresponding running law for Newton coupling is
\begin{equation}
 \label{eq:mod2G}
G(k)=\frac{B}{k^{\beta}}  \, .
\end{equation}
Here $A$ and $B$ are constants and the IR cutoff $k$ plays the role of the running scale. 

\subsection{Examples}

\label{examples}

\subsubsection{Vacuum space}

The simplest situation is that of the vacuum space with no matter, i.e. where $T^{\alpha \beta}=0$. For such a situation the Eq. (\ref{eq:cons2}) yields
\begin{equation}
 \label{eq:vacc}
g^{\alpha \beta} (\partial_{\alpha} \mu) \frac{d}{d \, \mu} (G(\mu) \rho_{\Lambda}(\mu)) = 0 \, .
\end{equation}
As in general the product $G(\mu) \rho_{\Lambda}(\mu)$ is dependent on $\mu$, the space-time variation of the scale $\mu$ must satisfy the condition
\begin{equation}
 \label{eq:scalevacc}
g^{\alpha \beta} (\partial_{\alpha} \mu) = 0 \, .
\end{equation}
As this condition is essentially a system of four equations with four unknowns and for all space-time points where the determinant of the $g^{\alpha \beta}$ matrix is nonvanishing, there is no space-time variation of $\mu$, i.e. $\partial_{\alpha} \mu = 0$ for all $\alpha$.

Strictly speaking, for vacuum spaces it is not possible to set the scale using our scale-setting procedure. However, the scale-setting procedure shows that there is no space-time dependence of the RG scale and that in vacuum spaces the parameters in action (\ref{eq:GRclass}) can be considered constant.

\subsubsection{Isotropic and homogeneous 3D space - ``cosmology''}

An isotropic and homogeneous 3D space, which is characteristic of our universe at large scales, is characterized by the Robertson-Walker metric
\begin{equation}
 \label{eq:RWmetric}
d s^2= d t^2 - a(t)^2 \left( \frac{d r^2}{1-k r^2} + r^2 d \theta^2 + r^2 \sin^2 \theta d \varphi^2 \right) \, .
\end{equation}
The four-vector of ideal fluid velocity in comoving coordinates is $u^0=1$, $u^i=0$. As the space is homogeneous and isotropic, the symmetry dictates that the RG scale $\mu$ can depend only on cosmic time, i.e., $\mu=\mu(x^0)=\mu(t)$. Using this information, one readily obtains that $u^{\alpha} \partial_{\alpha} \mu \neq 0$. From (\ref{eq:cons4}) one obtains 
\begin{equation}
 \label{eq:cosm}
G'(\mu)(\rho+\rho_{\Lambda}(\mu))+G(\mu) \rho_{\Lambda}'(\mu) = 0 \, .
\end{equation}
The analysis of Eq. (\ref{eq:cons3}) for all values of index $\beta$ reveals that (\ref{eq:cosm}) is the only scale-setting condition that one obtains for this space.
Indeed, the scale-setting condition already  obtained in a detailed analysis in \cite{miprd} is reproduced.

\subsubsection{Spherically symmetric static 3D space - ``star''}

In a spherically symmetric static 3D space the metric can be written down as \cite{weinbook}
\begin{equation}
 \label{eq:spher}
d s^2= B(r) d t^2-A(r) d r^2 - r^2 d \theta^2 - r^2 \sin^2 \theta d \varphi^2\, .
\end{equation}
The components of the ideal fluid four-velocity are $u^0 \neq 0$ and $u^i=0$. The spherical symmetry imposes that the RG scale should depend on the radial coordinate only, i.e. $\mu=\mu(x^1)=\mu(r)$. The quantity $u^{\alpha} \partial_{\alpha} \mu$ vanishes and (\ref{eq:cons4}) provides no useful information. The analysis of (\ref{eq:cons3}) for all values of index $\beta$ shows that in this case the following scale-setting condition is valid: 
\begin{equation}
 \label{eq:sphersett}
G'(\mu) (\rho_{\Lambda}(\mu)-p)+G(\mu) \rho_{\Lambda}'(\mu) = 0 \, .
\end{equation}

\subsubsection{Axisymmetric stationary 3D space - ``rotating galaxy''}

The axisymmetric stationary system in general has the metric of the form \cite{axi}
\begin{equation}
 \label{eq:axi}
d s^2= e^{2 \chi} d t^2 - e^{2 \psi} r^2 (d \varphi-\omega d t)^2 -e^{2 	\beta}(d r^2+d z^2)\, ,
\end{equation}
where the identification of coordinates is $x^0=t$, $x^1=r$, $x^2=\varphi$, and $x^3=z$ and $\chi$, $\psi$, and $\beta$ are functions of $r$ and $z$. 
The components of the ideal fluid four-velocity are $u^0 \neq 0$, $u^1=0$, $u^2 \neq 0$ and $u^3=0$. The symmetry of the system results in RG scale $\mu$ dependent on coordinates $r$ and $z$ only, i.e. $\mu=\mu(x^1,x^3)=\mu(r,z)$. In a axisymmetric stationary system  $u^{\alpha} \partial_{\alpha} \mu = 0$ and we obtain no scale-setting condition from (\ref{eq:cons4}). The full analysis of (\ref{eq:cons3}) on the other hand yields the scale-setting condition 
\begin{equation}
 \label{eq:axisett}
G'(\mu) (\rho_{\Lambda}(\mu)-p)+G(\mu) \rho_{\Lambda}'(\mu) = 0 \, .
\end{equation}
  
\subsubsection{Scale identification in astrophysical systems}

The scale-setting condition (\ref{eq:cosm}) arising in cosmology has been analyzed in detail in \cite{miprd} and it will not be further discussed here. It is intriguing that the scale-setting procedure  for spherically symmetric static systems and axisymmetric  stationary rotating systems yields the same scale-setting condition given by (\ref{eq:sphersett}) and (\ref{eq:axisett}). Whereas in cosmological setting the density of matter $\rho$ determines the running scale $\mu$, for the analyzed symmetric systems that may serve as a description of astrophysical objects the pressure $p$ determines the RG running scale. In particular, we may write (\ref{eq:sphersett}) and (\ref{eq:axisett}) as 
\begin{equation}
 \label{eq:fodp}
f(\mu) \equiv \rho_{\Lambda}(\mu) + \rho_{\Lambda}' (\mu) \frac{G(\mu)}{G'(\mu)} = p \, .
\end{equation}
The inversion of this expression then yields the functional form for the RG running scale
\begin{equation}
 \label{eq:fminus1p}
\mu=f^{-1}(p) \, .
\end{equation}

The application of this procedure to the first running model defined by (\ref{eq:mod1rho}) and (\ref{eq:mod1G}) yields the scale-setting condition
\begin{equation}
 \label{condmod1}
c_2 \mu^2 \left( 1 - d_2 +d_2 \ln \frac{\mu^2}{\mu_0^2} \right) = (c_0-p) d_2 \, .
\end{equation}
Under the assumptions of weak running of $G$, i.e., $d_2 \ll 1$ and that the CC energy density is negligible compared to the pressure, $ c_0 \ll p$ we obtain the expression
\begin{equation}
 \label{eq:exprmu} 
\mu^2=-\frac{d_2}{c_2} p \, .
\end{equation}

Similarly, for the second model defined by (\ref{eq:mod2rho}) and (\ref{eq:mod2G})  we obtain
\begin{equation}
 \label{eq:exprk}
k=\left(- \frac{ 8 \pi B \beta}{\alpha A} p \right)^{\frac{1}{\alpha+\beta}} \, .
\end{equation}

\section{Scale-setting in spherically symmetric systems}

\label{spherical}

For a spherically-symmetric system such as a star, it is in principle possible to establish an analytical connection between the running scale $\mu$ and the Newtonian potential $\phi$. The precise form of this analytical connection depends on the running laws for $G(\mu)$ and $\rho_{\Lambda}(\mu)$ underlying the model. 

Let us consider a spherically-symmetric object in which all quantities of interest have only radial dependence. The metric of space-time is given by (\ref{eq:spher}) and the hydrostatic equilibrium yields the Tolman-Oppenheimer-Volkoff equation \cite{weinbook}:

\begin{equation}
 \label{eq:hydro}
r^2 p' = - G_0 {\cal M}(r) (\rho+p) \left(1 + \frac{4 \pi G(\mu) r^3 (p-\rho_{\Lambda})}{G_0 {\cal M}(r)} \right) \left( 1-\frac{2 G_0 {\cal M}(r)}{r} \right)^{-1} \, ,
\end{equation}
  
where 
\begin{equation}
 \label{eq:calM}
{\cal M}(r)=\int_{0}^{r} 4 \pi \frac{G(\mu)}{G_0} (\rho+\rho_{\Lambda}(\mu)) r'^{2} d r' \, ,
\end{equation}
with the initial condition ${\cal M}(0)=0$.

In many astrophysical systems, except the most compact stars and galactic centers, the relativistic effects are not crucial for the explanation of their structure and dynamics and $\frac{2 G_0 {\cal M}(r)}{r} \ll 1$. The effect of $\rho_{\Lambda}(\mu)$ is very small in astrophysical systems and it can be safely taken that $\rho_{\Lambda}(\mu) \ll \rho$. Furthermore, in such systems the matter is nonrelativistic, which implies $p \ll \rho$. Finally, for the purposes of the identification of the running scale $\mu$ we can take that the running of $G$ can be neglected, i.e. $G(\mu) \simeq G_0$.  

In a spherically-symmetric system where the conditions put forward in the preceding paragraph apply equation (\ref{eq:hydro}) simplifies to \cite{weinbook}
\begin{equation}
 \label{eq:nonrel}
\frac{d}{dr} \left(\frac{r^2}{\rho(r)} \frac{dp}{dr} \right) = -4 \pi G_0 \rho(r) r^2 \, ,
\end{equation}
with an initial condition $p'(0)=0$.

Finally, we assume that the fluid is spatially isentropic \cite{weinbook} and that it obeys an equation of state in the form $p=p(\rho)$ which relates the energy density and the pressure of matter. Furthermore, in the case when the internal energy density of the matter fluid is proportional to its pressure and the energy density $\rho$ is dominated by the particle masses \cite{weinbook} the equation of state acquires the {\em polytropic} form
\begin{equation}
 \label{eq:eospoly}
p=K \rho^{\gamma} \, ,
\end{equation}
where $K$ and $\gamma$ are constants. 

For a spherically symmetric system the Newtonian potential $\phi$ satisfies the relation
\begin{equation}
 \label{eq:derpot}
\frac{d \phi}{d r}= G_0 \frac{{\cal M}(r)}{r^2} \, .  
\end{equation}
Combining (\ref{eq:nonrel}) and (\ref{eq:derpot}) one readily obtains the relation between $p$, $\rho$ and $\phi$:
\begin{equation}
 \label{eq:prhophi}
\frac{d p}{d r}=-\rho \frac{d \phi}{d r} \, .
\end{equation}
Taking into account the polytropic equation of state (\ref{eq:eospoly}) we arrive at the relation connecting the pressure and the Newtonian potential
\begin{equation}
 \label{eq:pvsphi}
p=\left[ \frac{\gamma-1}{\gamma} \frac{-\phi + C}{K^{1/\gamma}} \right]^\frac{\gamma}{\gamma-1}
\end{equation}
An especially interesting approximate solution is obtained in the regions where $|\phi| \gg C$ where
\begin{equation}
\label{eq:phiapprox}
p \sim (-\phi)^{\frac{\gamma}{\gamma-1}} \, . 
\end{equation}

Using this result, for the running model defined by (\ref{eq:mod1rho}) and (\ref{eq:mod1G}) we obtain
\begin{equation}
\label{eq:mudef}
\mu \sim (-\phi)^{\frac{\gamma}{2(\gamma-1)}}
\end{equation}
It should be remembered that this nontrivial result was obtained as a combination of the relation of the RG scale and the pressure (\ref{eq:fodp}), the condition of the hydrostatic equilibrium (\ref{eq:prhophi}) and the polytropic equation of state (\ref{eq:eospoly}).

Employing the notation $\alpha \equiv \frac{\gamma}{2(\gamma-1)}$ we obtain the Ansatz (\ref{eq:ansatz}). This result and its entire derivation also provides further insight into relation between the mass of the system $M$ and the exponent $\alpha$. Namely, in a more massive object we could expect that the matter equation of state is different than the matter equation of state in a less massive object. Therefore it is natural to expect that the parameter $\gamma$ in (\ref{eq:eospoly}) and consequently the parameter $\alpha$ depend on $M$.

For the running model given by (\ref{eq:mod2rho}) and (\ref{eq:mod2G}) the running scale becomes
\begin{equation}
\label{eq:kdef}
k \sim (-\phi)^{\frac{\gamma}{(\gamma-1)(\alpha+\beta)}} \, .
\end{equation}

Finally, the analysis of the RG scale-setting in spherically symmetric systems presented in this section provides additional insight into the relation of the exponent $\gamma$ from (\ref{eq:eospoly}) (or consequently the exponent $\alpha$ from (\ref{eq:ansatz})) and the mass $M$ of the system. Using the redefinitions \cite{weinbook}
\begin{equation}
 \label{eq:relab}
r = \left( \frac{K \gamma}{4 \pi G_0 (\gamma-1)} \right)^{1/2} \rho(0)^{(\gamma-2)/\gamma} \xi \, , \;\;\; \rho= \rho(0) \theta^{1/(\gamma-1)} \, , 
\end{equation}
the equation (\ref{eq:nonrel})  can now be represented as Lane-Emden equation
\begin{equation}
 \label{eq:LEeq}
\frac{1}{\xi^2} \frac{d}{d\xi} \left( \xi^2 \frac{d \theta}{d \xi} \right) + \theta^{1/(\gamma-1)}=0
\end{equation}
and the corresponding initial conditions are $\theta(0)=1$ and $\theta'(0)=0$.

Following the analysis of \cite{weinbook}, one readily obtains the relation between the mass $M$ and radius $R$ of the star and the parameter $\gamma$:
\begin{equation}
 \label{eq:MRgamma}
M= 4 \pi R^{(3 \gamma-4)/(\gamma-2)} \left( \frac{K \gamma}{4 \pi G_0 (\gamma-1)} \right)^{-1/(\gamma-2)} \xi_1^{-(3 \gamma-4)/(\gamma-2)} (-\xi_1^2 \theta'(\xi_1) )\, .
\end{equation}
Here $\theta(\xi_1)=0$ and $\xi_1$ corresponds to the radius $R$ of the spherical system (star).

This equation provides a direct connection between the mass of the system $M$, its size $R$ and the parameter of its equation of state $\gamma$ (and thus also $\alpha$). It is reasonable to expect that a similar relation between the mass, the size and the equation of state parameters might hold for other astrophysical objects such as galaxies. In \cite{galrot} the importance of the relation between the mass $M$ and the parameter $\alpha$ was especially stressed. For a spherically symmetric system, the expression (\ref{eq:MRgamma}) represents such a relation, but the size of the system also has to be taken into account.

\section{Discussion}

The identification of the running scale performed in section \ref{spherical} primarily serves as an analytically tractable demonstration of the scale-setting procedure in astrophysical systems. However, it is striking to see that under simple assumptions it is possible to systematically obtain the identification of the running scale introduced as an Ansatz in \cite{galrot}. This fact that this Ansatz reproduces the galactic rotation curves without need for dark matter distribution deserves a more detailed discussion.

The analysis presented in section \ref{spherical} clearly represents a justification for the Ansatz of \cite{galrot}, but it also requires considerable extension and refinement to be applied to the analysis of empirical galactic rotation data. First important issue is the interplay of rotation and pressure in providing stability of a rotating galaxy. Another important point is the equation of state of matter in the galaxy (which does not have to be polytropic) as well as to which extent the matter can be treated as an ideal fluid. For a realistic galaxy it is expected that considerable numerical work is required. The type of numerical analysis performed in \cite{galrot} should be supplemented by the numerical determination of the scale from the scale-setting condition (\ref{eq:sphersett}).
 
A more theoretical question is whether the scale-setting procedure results in a reasonable identification of the running scale for any type of running laws, such as (\ref{eq:mod1rho}) and (\ref{eq:mod1G}). Although this issue requires additional investigation, from (\ref{eq:mudef}) it is possible to see that the model coefficients $c_2$ and $d_2$ must have opposite signs for positive pressure. It is expected that for a well justified running laws the scale-setting procedure would yield meaningful results for the RG running scale.

%\begin{figure}
%\centerline{\resizebox{0.6\textwidth}{!}{\includegraphics{lamnega-2.5.eps}}}
%\caption{\label{fig:lnega} The evolution of $h=H^2/H_X^2$ as a function of the scale factor for different values of the exponent $\alpha$. The value of $\alpha$ strongly influences the asymptotic value of $h$ at large $a$, whereas the behavior at small $a$ is not affected by $\alpha$. The values of the parameters used are $\lambda=-1000$, $\xi=0.01$ and $w=-0.9$.}
%\end{figure}

\section{Conclusions}

\label{conc}

The problem of systematic determination of the running scale in RGGR deserves attention for at least two distinct reasons. First, a systematic procedure for the determination of the running scale is needed to reduce model dependence in various approaches in RGGR. The second reason is that recent developments in cosmology (see e.g. \cite{mijcap}) and understanding of galactic rotation curves \cite{galrot} in the framework of RGGR use {\em ad hoc} choices for the scale and a systematic validation of these choices is needed. In this paper we present a systematic scale-setting procedure for RGGR and elaborate it for systems with various types of symmetry.
In particular, for a spherically symmetric system with matter obeying a polytropic equation of state we reproduce the choice of scale capable of explaining the galaxy rotation curves in the framework of RGGR, using the developed scale-setting procedure. This result reinforces the approach to the galaxy rotation curves in the framework of RGGR. Although a numerically challenging analysis in a realistic galactic setting seems necessary, a prospect of explaining galaxy rotation curves in RGGR using a systematic procedure seems rewarding enough to engage into such an endeavor.

\vspace{2cm}

{\bf Acknowledgements.} The authors would like to thank Ilya Shapiro, Branko Guberina, Raul Horvat and Neven Bili\'{c}  for useful comments on the manuscript. This work was
supported by the Ministry of Education, Science and Sports of the Republic of Croatia 
under the contract No. 098-0982930-2864.

\end{document}